\documentstyle[pre,aps,preprint,epsfig,psfrag]{revtex}
\def\subequations{\begin{mathletters}}
\def\endsubequations{\end{mathletters}}
\def\gp{G'(\omega)}
\def\gpp{G''(\omega)}
\def\g*{G^*(\omega)}
\def\o{\omega}
\def\gdot{\dot{\gamma}}
\def\gs{\dot{\gamma_s}}
\def\etal{{\it et al.}}
\begin{document}
\bibliographystyle{unsrt}


\title{Structure and Rheology of the Defect-gel States of Pure
and Particle-dispersed Lyotropic Lamellar Phases}
\author{Geetha Basappa$^1$, Suneel$^2$, V. Kumaran$^2$,
Prabhu R Nott$^2$, Sriram Ramaswamy$^1$, V.M. Naik$^3$, and Deeleep
Rout$^3$}
\address{$^1$ Centre for Condensed Matter Theory,
Department of Physics, $^2$Department of Chemical
Engineering, Indian Institute of Science, Bangalore 560012 INDIA,
$^3$Unilever Research India, 64 Main Road, Whitefield, Bangalore 560 066
INDIA\\ E-mail: lcrheol@chemeng.iisc.ernet.in}
\maketitle
\begin{abstract}
We present
important new results from light-microscopy and rheometry on a
moderately concentrated lyotropic smectic, with and without particulate
additives. Shear-treatment aligns the phase
rapidly, except for a striking network of oily-streak defects, which
anneals out much more slowly.
If spherical particles several microns in diameter are dispersed in
the lamellar medium, part of the defect network persists under
shear-treatment, its nodes
anchored on the particles. The sample as prepared has substantial
storage and loss moduli, both of which decrease steadily under
shear-treatment. Adding particles enhances the moduli and
retards their decay under shear. The data for the frequency-dependent storage
modulus after various durations of shear-treatment can be scaled
to collapse onto a single curve.
The elasticity and dissipation in these samples thus arises mainly from
the defect network, not directly from the smectic elasticity and hydrodynamics.
\end{abstract}

\pacs{PACS no.: 61.30Jf, 61.30Eb, 83.70}

\section{Introduction and Results}
\label{intro}
Many useful materials arising in the domain of chemical engineering are
liquid-crystalline, frequently lamellar. They are generally processed
under conditions of shear flow, and often structured by the addition of
particles \cite{shouche} . The complex dynamic modulus $\g*=\gp + i
\gpp$ of these materials as a function of the
 angular frequency $\o$  (=2$\pi f$) presents many puzzles. For example, the lamellar
L$_{\alpha}$ phase, although in principle a one-dimensional stack of
two-dimensional fluid layers and hence incapable, in the ideal, perfectly
ordered state, of sustaining a static
shear stress, generally displays \cite{larson3} in practice a modest
storage modulus $\gp$ at small $\o$. In addition, it has an anomalously
large low-frequency loss modulus $\gpp$. Since many soft solids,
lamellar or otherwise, share these features\cite{cates}, it is clear
that rheometry alone cannot discriminate between various models for these
materials.

Since the observed rheometric properties of lamellar phases (for some
recent studies see \cite{diat,larson2,meyer}) are presumably
a consequence of topological defects and textures, one must first
correlate the rheometry to direct visual observations. One could
then hope to build a theory in two stages, by explaining first how
the defects produce the rheology, and second why flow or sample
preparation produces the defects to begin with.
A major motivation for the present study, in particular, is the
observation \cite{shouche} of a large enhancement of the shear
modulus of L$_{\beta}$ gels upon the addition of a very small
concentration of particles.
Accordingly, this paper reports parallel studies of the viscoelastic
 properties and defect structure of the lamellar
 phase, and
the effect thereon of shear-treatment and particulate additives.
Our main results are as follows:
\begin{itemize}
\item {\it Pure lamellar phase without shear-treatment}: The
L$_{\alpha}$ samples have $\gp \sim 10^3$ Pa and $\gpp \sim 10^2$ Pa at
$\o = 1$ rad/sec. Both $\gp$ and $\gpp$ depend only weakly on $\omega$,
with $\gp$ nearly flat. Between crossed polars, the system appears dense
with defects.
\item {\it Shear-treatment without added particles}: If the system is
subjected to steady shear at a low rate $\gs$ ( 1 to 50
s$^{-1}$) for a specified duration $t_s$, the small-amplitude
frequency-dependent moduli measured after switching off the steady shear
depend strongly on both $\gs$ and $t_s$. In general, the moduli thus
measured decrease with increasing $t_s$ for fixed $\gs$, but appear
roughly to level off at a value which is an increasing function of
$\gs$. The $\gp,\,\gpp$ reported here seem to depend weakly on the
shear history. Observations between crossed polars show that the
shear-treated samples consist of macroscopically well-oriented lamellae
parallel to the applied velocity, and normal to the velocity gradient,
threaded by a rather beautiful network of oily-streak defects
\cite{kleman,klemanbook} (see
section \ref{defnopart} for background on these defects), as
seen recently in cholesterics \cite{zap}. It is
thus clear that the dominant contribution to the rheology comes from
this defect network in an otherwise well-oriented phase \cite{zap}, not
from a polydomain averaging of the smectic elasticity, as suggested
by \cite{kawa}.
\item {\it Effect of particulate additives}: Suspending micron size
particles had a dramatic effect on the structure and rheology. Prior to
shear treatment both $G '$ and $G ''$ are appreciably larger. Under
shear-treatment, the oily-streak network survives much longer, with the
spheres located at its nodes, and the measured $G'$ and $G ''$ persist
longer. This picture of a particle-stabilised defect network is
consistent with the observations of \cite{zap} on colloid-doped
cholesterics. It is worth adding that careful observations under the
polarising microscope reveal unambiguously that the alignment at the
particle surface is homeotropic.
\end{itemize}

The relevance of our work to that of Shouche \etal \cite{shouche}
should be emphasised here. It was conjectured in \cite{shouche} that
the enhancement of elasticity when particles are added to an L$_{\beta}$
gel arose from interparticle bridges formed by surfactant molecules.
Our study suggests strongly that oily-streak defects, not surfactant
molecules, form the bridges, and we are now turning our attention to
L$_{\beta}$ phases to see if this is so.

The remainder of this paper is organised as follows. In section II we
specify the samples used and describe our experimental setup. In section
III we present in detail our visual observations of the defect structure
and its evolution under shear, with and without added particles. The
steady-shear and small-amplitude rheometry of these systems are
discussed in Section IV. We close in section V with a tentative
interpretation and analysis of our results.

\section{Experimental}
\label{exp}
\subsection{Sample and sample preparation}
\label{prep}

The lamellar (L$_{\alpha}$) sample is a commercially available (Galaxy)
anionic surfactant, sodium dodecyl ether sulphate or SLES (73.2 wt.\%)
in water with a substantial but unspecified concentration of ionic
impurities.  For some of the experiments we have diluted this sample with
distilled water and homogenized the mixture in the oven at $\sim$
70$^\circ$C for a few hours. We have dispersed 9.55$\pm$0.44 $\mu$m
polystyrene + 30\% polybutylmethacrylate (initially suspended in water, Bangs Lab) spheres at
nominal volume fractions of $\sim$ 0.5 \% in the L$_{\alpha}$ sample.
The mixtures were made by manually mixing in the particles with a glass
rod or spatula. The samples were centrifuged briefly to remove air
bubbles.

\subsection{Imaging setup}
\label{imsetup}
 Our experimental studies of the flow properties of particle laden
lyotropic smectic liquid crystals are in two parts. While the
rheometry is carried out in a commercial rheometer (details below),
the visualisation of the defect structure under flow is done in
a custom-built flow cell. The design is based on one
available in the literature \cite{larson}, and was executed by Holmarc
Instruments.
\begin{figure}
\begin{center}
\epsfig{file=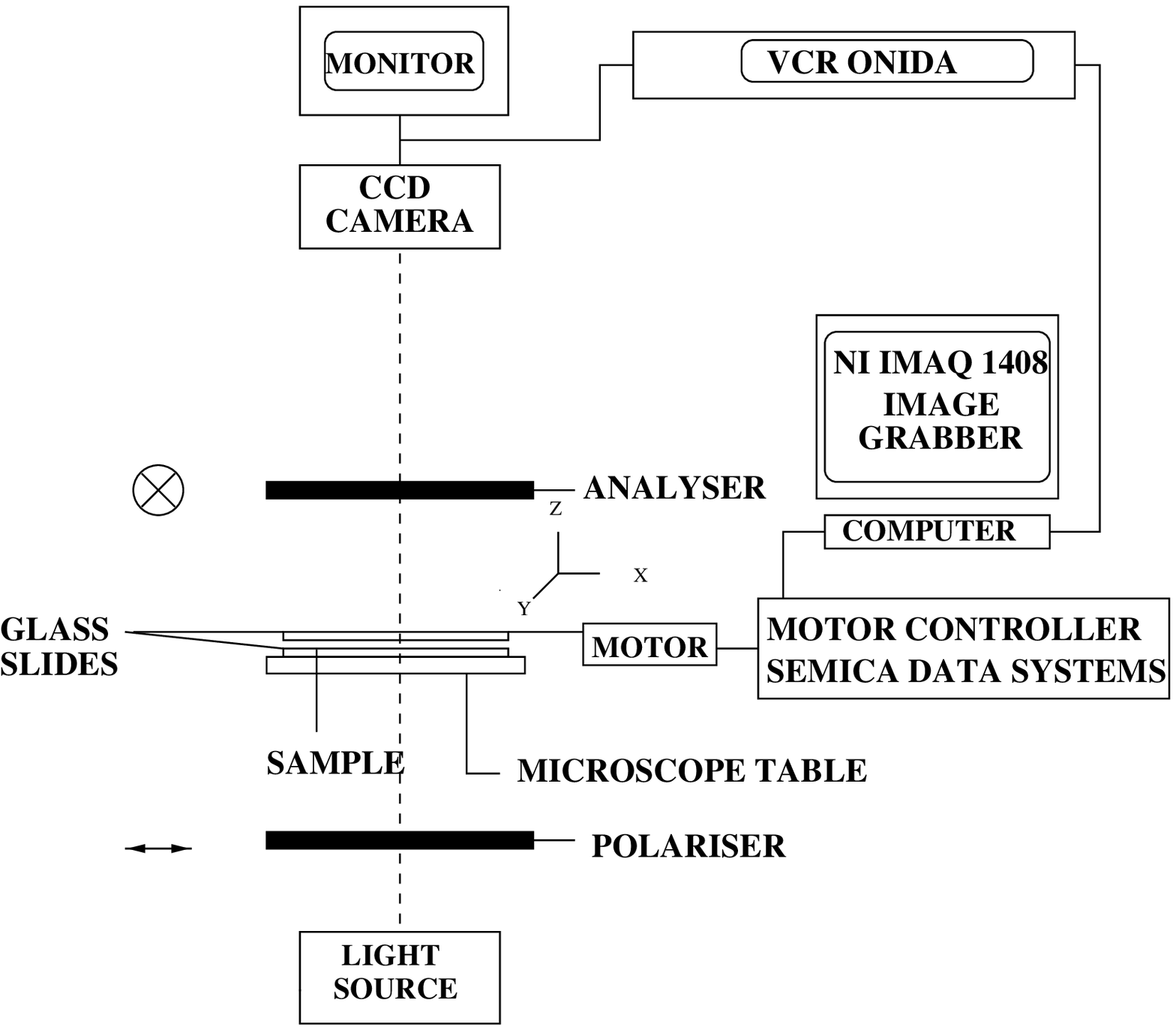,width=11cm,bbllx=0,bblly=0,bburx=477,bbury=415}
\end{center}
\vspace{1cm}
\caption{Schematic diagram of the rheooptics setup.}
\label{fig:imagesetup}
\end{figure}

Two horizontal glass plates (Borosilicate, BK7) lying parallel to the
$xy$ plane form the walls of the channel, and the sample is sheared
between these plates, providing a linear (plane Couette) flow with
velocity along $x$ and gradient along $z$. The stationary lower glass
plate is fixed to a metallic plate mounted on two micrometer screws,
which are used to ensure that the lower glass plate is parallel to the
upper plate. The upper plate is attached to the moving arm of a
translation stage, and the vertical ($z$) separation between the plates
adjusted using a micrometer (least count = 10$\,\mu$m) fixed to the
moving arm of the stage. The moving arm is mechanically linked to a
linear stepper motor via a lead screw (Holmarc Instruments), allowing a
minimum step size 0.005$\,$mm. The motor is controlled by a computer
using an ISA interface card (Semica Data Systems). This translation
stage is mounted on the microscope stage as shown in Fig.
\ref{fig:imagesetup}.

The sample is viewed through crossed polars using a Nikon Optiphot2-Pol
polarising microscope fitted with a CCD camera (Sony TK-S300) whose
output goes simultaneously to a video recorder and a computer for image
analysis (NI IMAQ 1408 image grabbing card). The line of sight is in the
direction of the velocity gradient. Note that a sample aligned
homeotropically, i.e., with layers parallel to the plates, should appear
dark through crossed polars.

The glass slides were cleaned thoroughly before the sample was loaded
and were not pretreated for any preferential alignment. The rheooptic
experiments were conducted at room temperature (25--28$^{\circ}$C).

Our rheometric studies have been carried out on a Rheolyst AR1000N (TA
Instruments) stress contolled rheometer. For all the experiments quoted
in this paper we have used a 4$\,$cm parallel plate geometry and the
temperature was maintained at 25 $^{\circ}$C.
\section{Defect structure of the lamellar phase, and its
evolution under shear}
\label{defects}
\subsection{Defect network of the lamellar phase without particles}
\label{defnopart}

\begin{figure}
\begin{center}
\epsfig{file=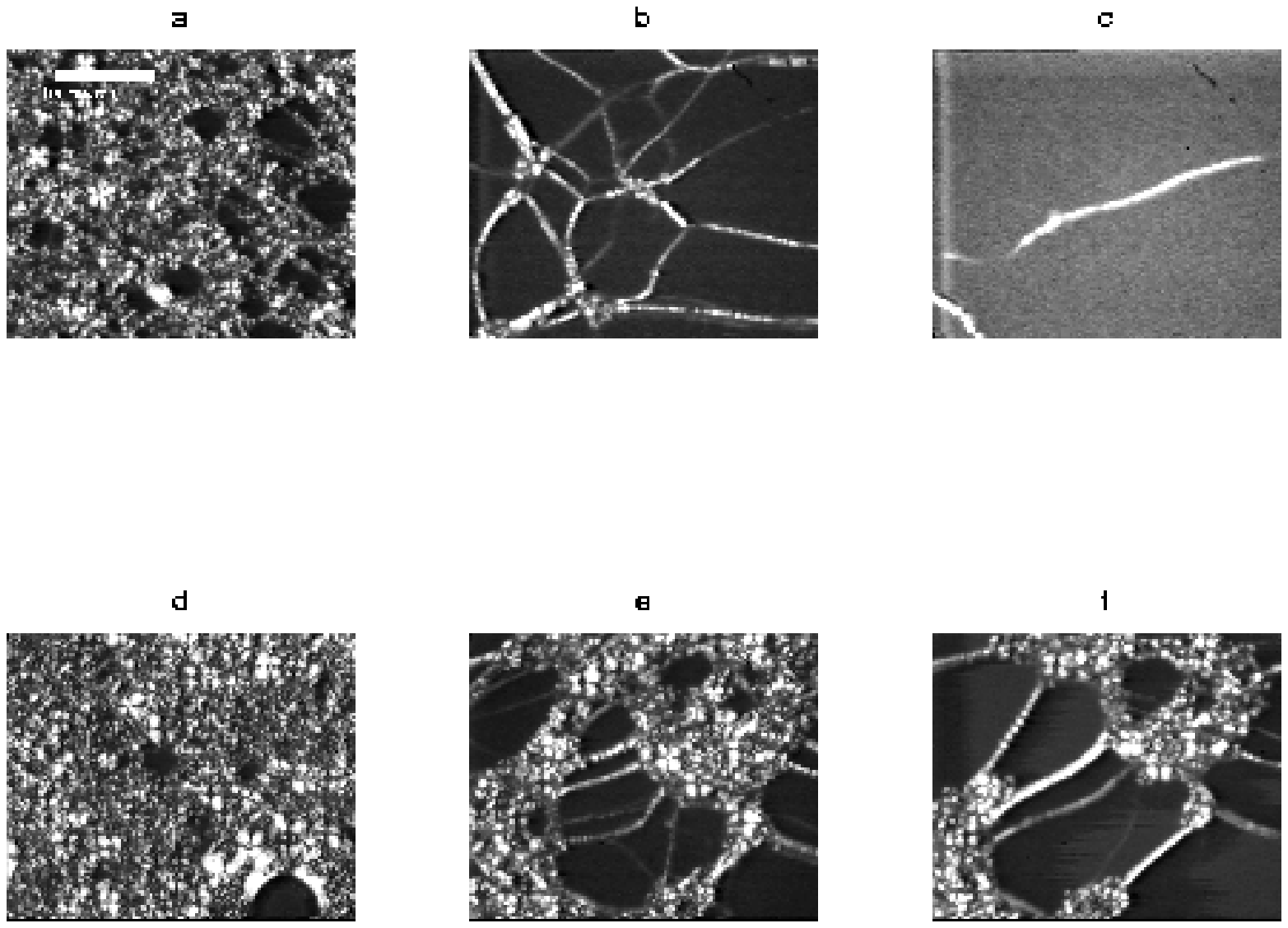,width=14cm}
\end{center}
\caption{Images of the lamellar liquid crystal (concentration of SLES=70w/w \%) with and without particles, as a function of shear time ($t_s$), $\gs$=2 s$^{-1}$. Without particles,
$t_s$=(a) 0, (b) 140 and (c) 350 sec. With particles (diameter 9.55
$\mu$m): $t_s$= (d) 0, (e) 140 and (f)= 350 sec. The magnification is
$\times$ 100. The white scale bar in (a) is 100 $\mu$m. The particles
are aggregated at the network nodes in (d) and (f). The dark patch at
the lower right hand corner of (d) is an air bubble.}
\label{fig:images}
\end{figure}

The sample when loaded presents a bright appearance because it is full
of defects. The texture is that of a typical lyotropic lamellar liquid
crystal. We then subject it to large-amplitude oscillatory shear of
frequency 0.1$\,$Hz and end-to-end amplitude 1$\,$mm. The sample
thickness $d$ is 100$\,\mu$m for the case we are reporting in this
paper, with a triangle-wave strain, so that in each phase of its motion
there is a constant shear-rate $\gdot = 2$ s$^{-1}$. The
structure observed between crossed polars evolves steadily from its
initial highly defect-ridden state, Fig. \ref{fig:images}(a), to one
with large dark (and hence homeotropic) regions, with a striking,
sparse, sample-spanning network of linear oily-streak \cite{kleman}
defects (Fig. \ref{fig:images}b), and finally to an almost totally
homeotropically oriented lamellar state with a weak background of
defects (Fig. \ref{fig:images}c). That the defects span the sample is
clear from the video because different portions of a typical defect line
move at different speeds under shear, indicating that they are at
different depths. This network structure is very similar in appearance
to that seen in \cite{zap}, and the behaviours of the two systems are
broadly alike.

We remind the reader that oily streaks \cite{klemanbook} are formed in
lamellar phases upon the close approach of a pair of
parallel dislocation lines with Burgers vectors with magnitude
$b_1$ and $b_2$ and opposite sign. Instead of annihilating partly
to yield a simple dislocation with Burgers vector of magnitude
$|b_1 - b_2|$, they display a complex internal structure, whose
nature depends on material properties such as the splay and saddle-splay
rigidity moduli of the lamellae. Oily streaks have been
discussed in detail in \cite{schn} and \cite{kleman}, and the latter paper
provides an explanation of the variety of inner structures observed
in these defects. The two main possibilities considered are: (i) the
dislocation lines retain a core structure consisting of a pair of
disclination lines, with modulations runing along their length,
and (ii) they nucleate an array of focal conic domains if the
saddle-splay modulus favours large negative Gaussian curvature.
Our system shows focal domains, and we will therefore use the model
of \cite{kleman} to estimate the line energy of the streaks in
section \ref{summary}.

The following important differences between our
experiments and those of \cite{zap} on cholesteric liquid crystals
remain to be understood: (i) The defect network in our study, unlike
that in \cite{zap}, shows no sign of coarsening in the absence of shear
treatment. Indeed, we observe no spontaneous annealing of the initially
defect-ridden structure unless sheared. Possibly the much larger ratio
of sample thickness to layer spacing in our study is responsible for
this difference; the absence of surface treatment of our plates could
also play a role. (ii) Our network under shear evolves mainly by the
{\em thinning} and eventual {\em disappearance} of lines; we almost never
\footnote{Our video footage contains one possible candidate for a
detachment-and-retraction event.} observe the {\em detachment} and {\em
retraction} events reported in \cite{zap}. It remains unclear whether
detachment events are taking place at scales unresolved by our imaging,
but thinning is clearly an important component of the process.
(iii) The terminal defect density is higher if the imposed shear-rate
is higher. (iv) The
defect line segments are not as straight as those in \cite{zap} appear
to be, and flex rather than rotate rigidly when sheared. This could have
implications for the elasticity of the network. (v) As the sample is
continuously sheared, regions which had initially become homeotropic
display the onset of a square grid like defect structure. A
similar type of shear induced defect structure has been seen by Larson
\etal\cite{larson2} in a thermotropic liquid crystal. This is due to
disruption of the monodomains in the form of undulation instabilities
\cite{os}. Lastly, we do not see, over the range of shear rate
examined, the multilamellar vesicles (``onions'' \cite{diat}) found in
shear studies of more dilute lamellar phases.

\vspace{1cm}

\subsection{The effect of suspended particles on the defect network}
\label{defpart}
The effect of a small concentration (nominally 0.5 \%) of polystyrene +
30\% polybutylmethacrylate spheres of diameter 9.55 $\mu$m on the
behaviour of the defect network was quite dramatic. Note first that the
initial configuration shows a clearly homeotropic anchoring of the
lamellar phase onto the particles. This can be seen explicitly, by
noting the colour variation as a function of angle around the particle
(colour pictures will be sent by the authors upon request, and may be
viewed at http://144.16.75.130/lcrheol/) when it is viewed with a
$\lambda$ (530$\,$nm) plate inserted with its vibration directions at
45$^\circ$ with respect to the extinction position of the crossed
polarisers, and using the colour charts provided in standard books on
polarising microscopy, e.g. \cite{hart}. Before the application of
shear, the particles appear to be well dispersed in the sample.

If we shear at 2 s$^{-1}$ as in the undoped case an oily streak network
appears. Now we find that the network after 140 seconds (Fig.\
\ref{fig:images}e), is much denser than in Fig.\ \ref{fig:images}(b)
with particles aggregated at its nodes. After 350 seconds of shearing
the network is still substantially present (Fig.\ \ref{fig:images}f),
suggesting that the particles are {\it stabilising} it against dissolution. As
the sample is being sheared there is a tendency for the particles to
aggregate and at much longer times, $\sim $900$\,$s, the defect
structure anneals out leaving an almost perfectly aligned sample in
particle-free regions. Similar behaviour was observed for a system doped
with 19$\,\mu$m silica particles. The detailed mechanism for the decay
of the network is unclear, as remarked above, and so therefore is the
reason for its stabilisation. We will, however, offer some speculations
on this subject below.

\section{Viscoelastic properties of the sheared lamellar phase,
and the effect thereon of particulate additives}
\label{viscoel}

\subsection{Shear-treatment and rheometry without particles}
\label{rheonopart}
Our aim is to correlate the visualisation studies of section III with
the mechanical properties of the lamellar phase. Accordingly, we try to
reproduce as far as possible the shear-treatment conditions of section
III and study its effects on the rheometry. The rheometry was carried
out in a parallel-plate geometry\footnote{While this geometry has a
nonuniform shear rate, we prefer it to the (viscometric) cone-and-plate
geometry because the wedge-shaped cross-section of the latter generates
a tilt grain boundary. In addition, there is the danger of particles
getting stuck in the narrow gap near the cone center.} with a gap of
100$\,\mu$m and plate diameter 4$\,$cm. $\gs$, the shear-rate of
shear-treatment referred to in all our rheometric studies, is
defined at the outer rim of the parallel-plate geometry.

\begin{figure}
\begin{center}
\psfrag{xlabel}{{\small $t_s(sec)$}}
\psfrag{ylabel}{$G'$(Pa)}
\epsfig{file=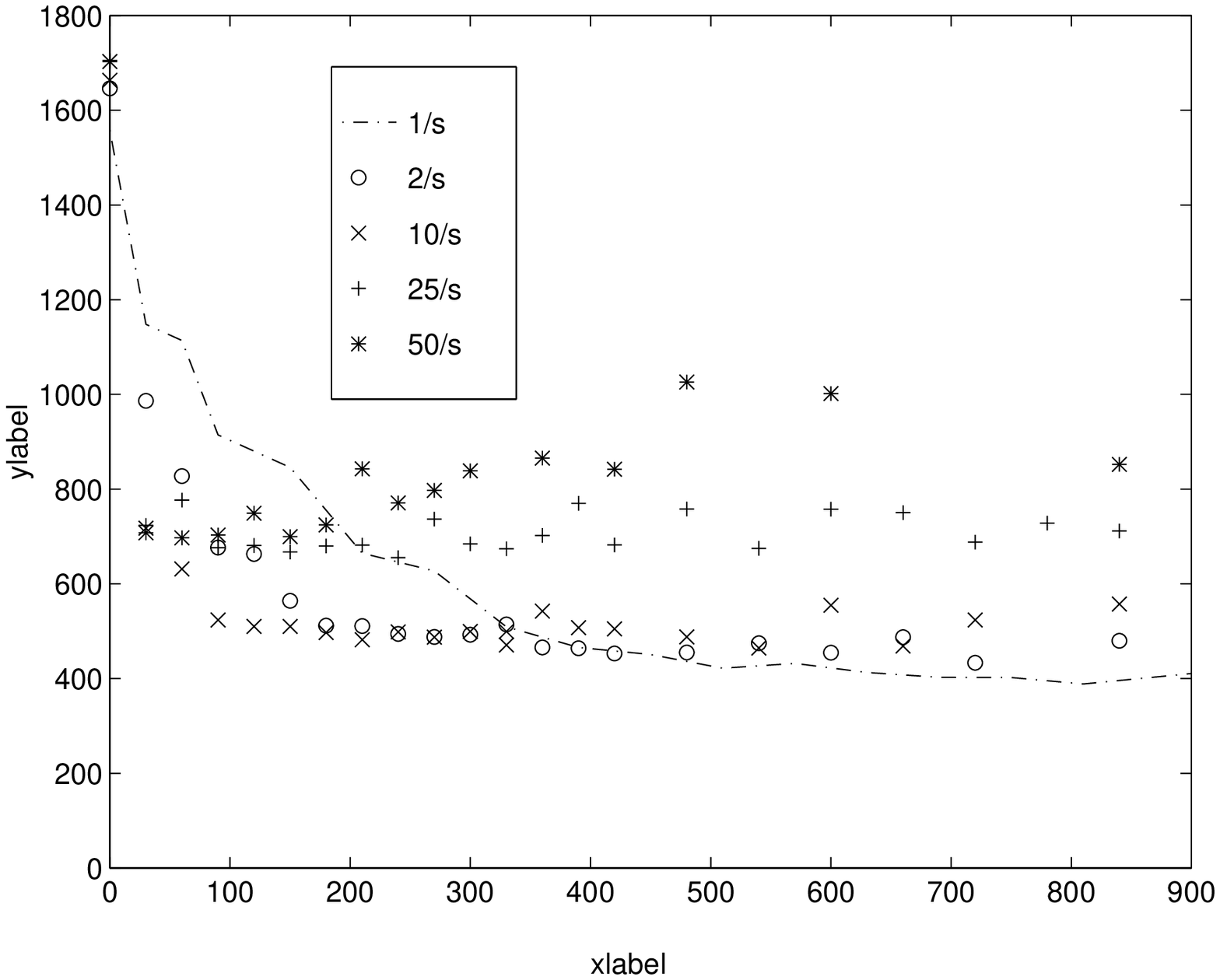,width=11cm,bbllx=90,bblly=246,
bburx =540,bbury=560}
\end{center}
\vspace{1cm}
\caption{Variation of $G'$ as a function of shear time $t_s$ for various shear rates.
Frequency, $f$=1Hz ($\o$=2$\pi$ rad/sec), T=25 $^{\circ}$C.  }
\label{fig:rheo1}
\end{figure}

\begin{figure}
\begin{center}
\psfrag{xlabel}{$t_s(sec)$}
\psfrag{ylabel}{$G''$(Pa)}
\epsfig{file=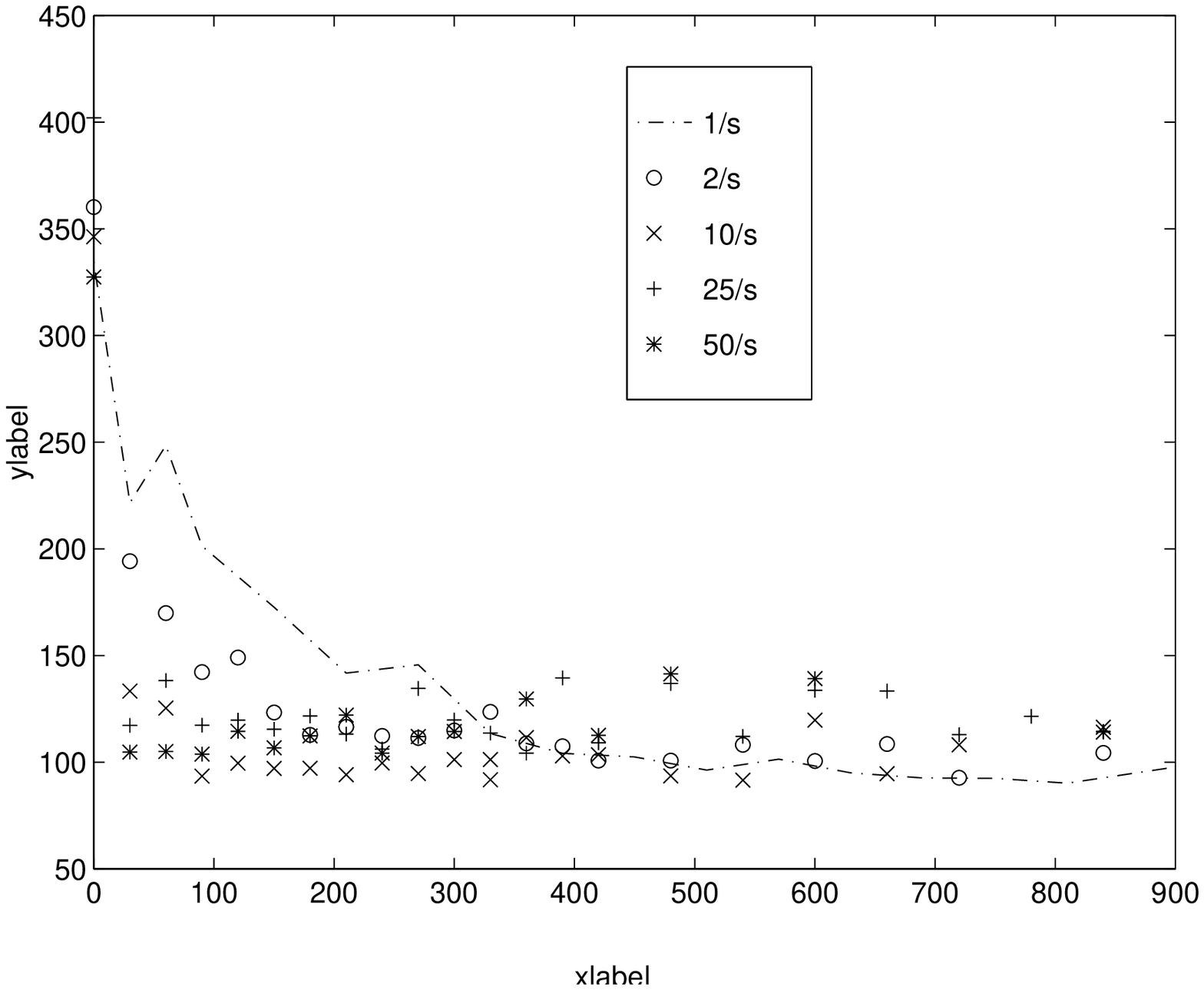,width=11cm,bbllx=90,bblly=246,
bburx =540,bbury=560}
\end{center}
\vspace{1.2cm}
\caption{Variation of $G''$ as a function of shear time $t_s$ for various shear rates.
 Frequency, $f$=1Hz, T=25 $^{\circ}$C.  }
\label{fig:rheo1g2}
\end{figure}

\begin{figure}
\begin{center}
\psfrag{xlabel}{$\omega$ (rad/sec)}
\psfrag{ylabel}{$G'$ (Pa)}
\epsfig{file=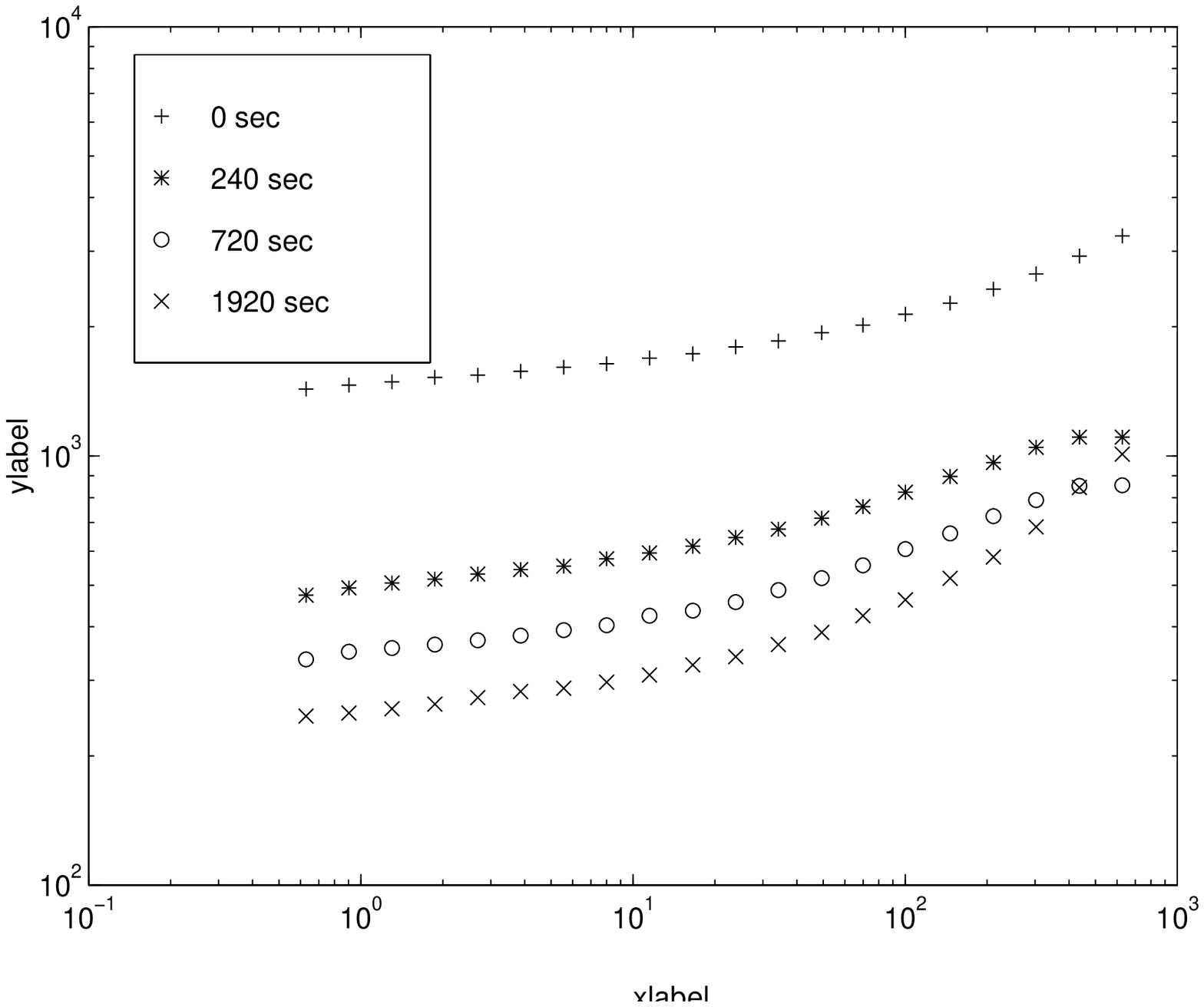,width=11cm,bbllx=90,bblly=246,bburx =540,bbury=560}
\end{center}
\vspace{1cm}
\caption{$\gp$ at different shear times $t_s$ for $\gs$=1s$^{-1}$, T=25$^\circ$C.
 }
\label{fig:rheo2}
\end{figure}

\begin{figure}
\begin{center}
\psfrag{xlabel}{$\o$(rad/sec)}
\psfrag{ylabel}{$G''$(Pa)}
\epsfig{file=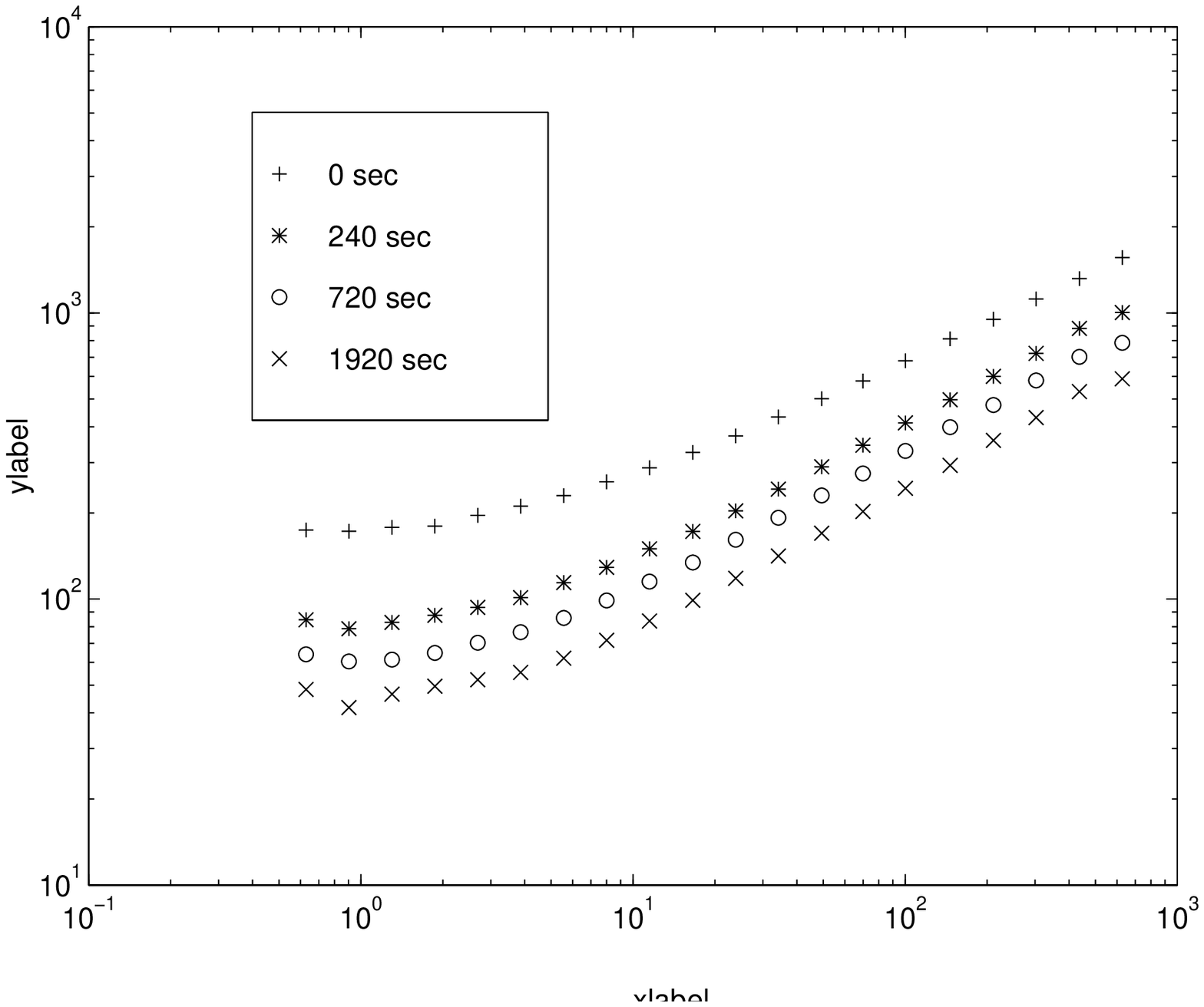,width=11cm,bbllx=90,bblly=246,bburx =540,bbury=560}
\end{center}
\vspace{1cm}
\caption{$\gpp$ at different shear times $t_s$ for $\gs$=1s$^{-1}$, T=25$^\circ$C.
 }
\label{fig:rheo2g2}
\end{figure}

\begin{figure}
\psfrag{xlabel}{log($ \o/\o_o$)}
\psfrag{ylabel}{log$(G'/G'_o)$}
\begin{center}
\epsfig{file=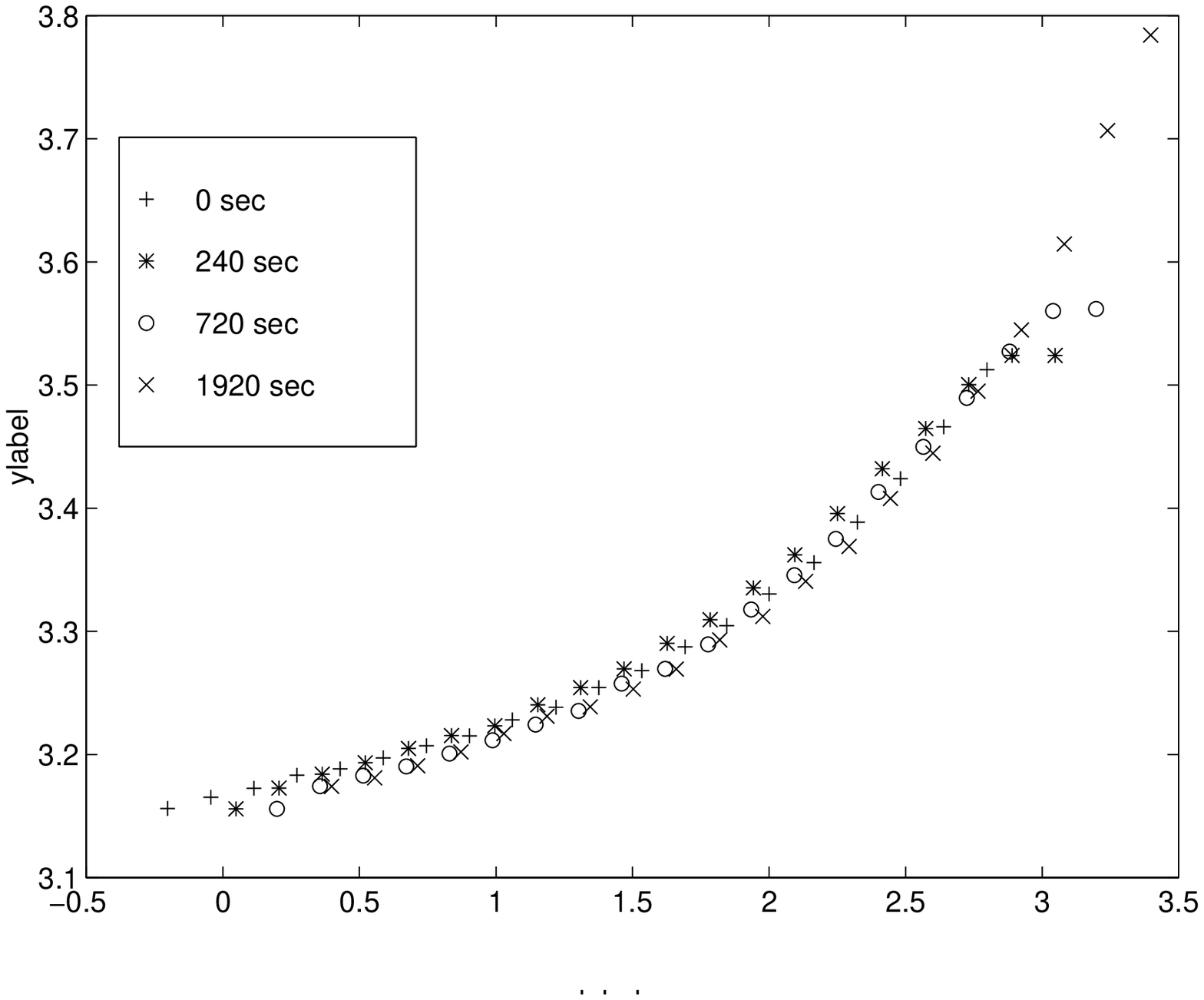,width=11cm,bbllx=90,bblly=246,bburx =540,bbury=560}
\end{center}
\vspace{1cm}
\caption{The $\gp$ data of Fig. \ref{fig:rheo2} for various shear
times collapsed onto the zero shear treatment ($t_s=0$) curve.
 }
\label{fig:rheo3}
\end{figure}

\begin{figure}
\begin{center}
\epsfig{file=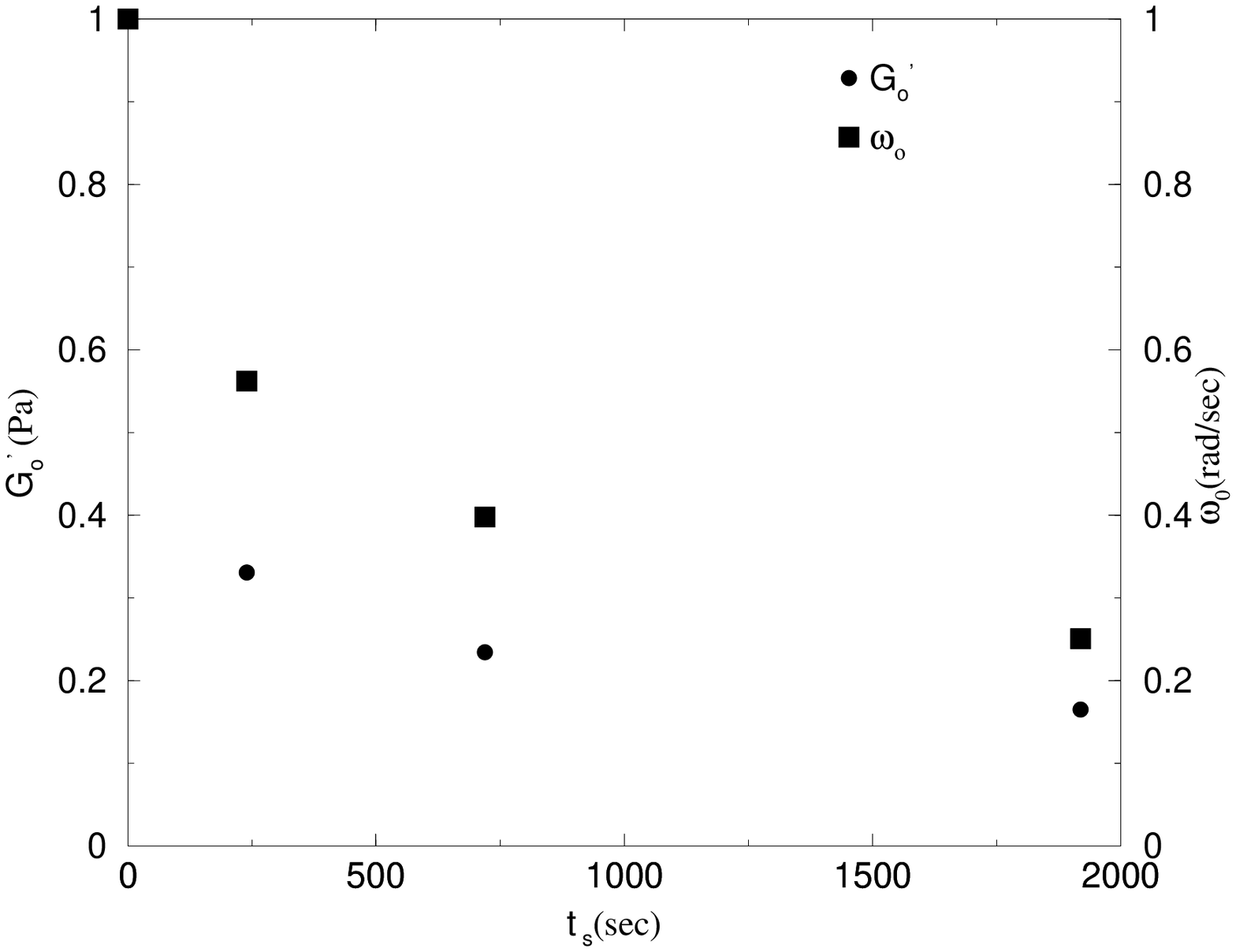,width=10cm}
\end{center}
\caption{Variation of the scale factors $G'_o$ and $\omega_o$ as functions
of shear time.
 }
\label{fig:go}
\end{figure}

\begin{figure}
\begin{center}
\psfrag{ylabel}{$t_s$(sec)}
\psfrag{xlabel}{$G'$(Pa)}
\epsfig{file=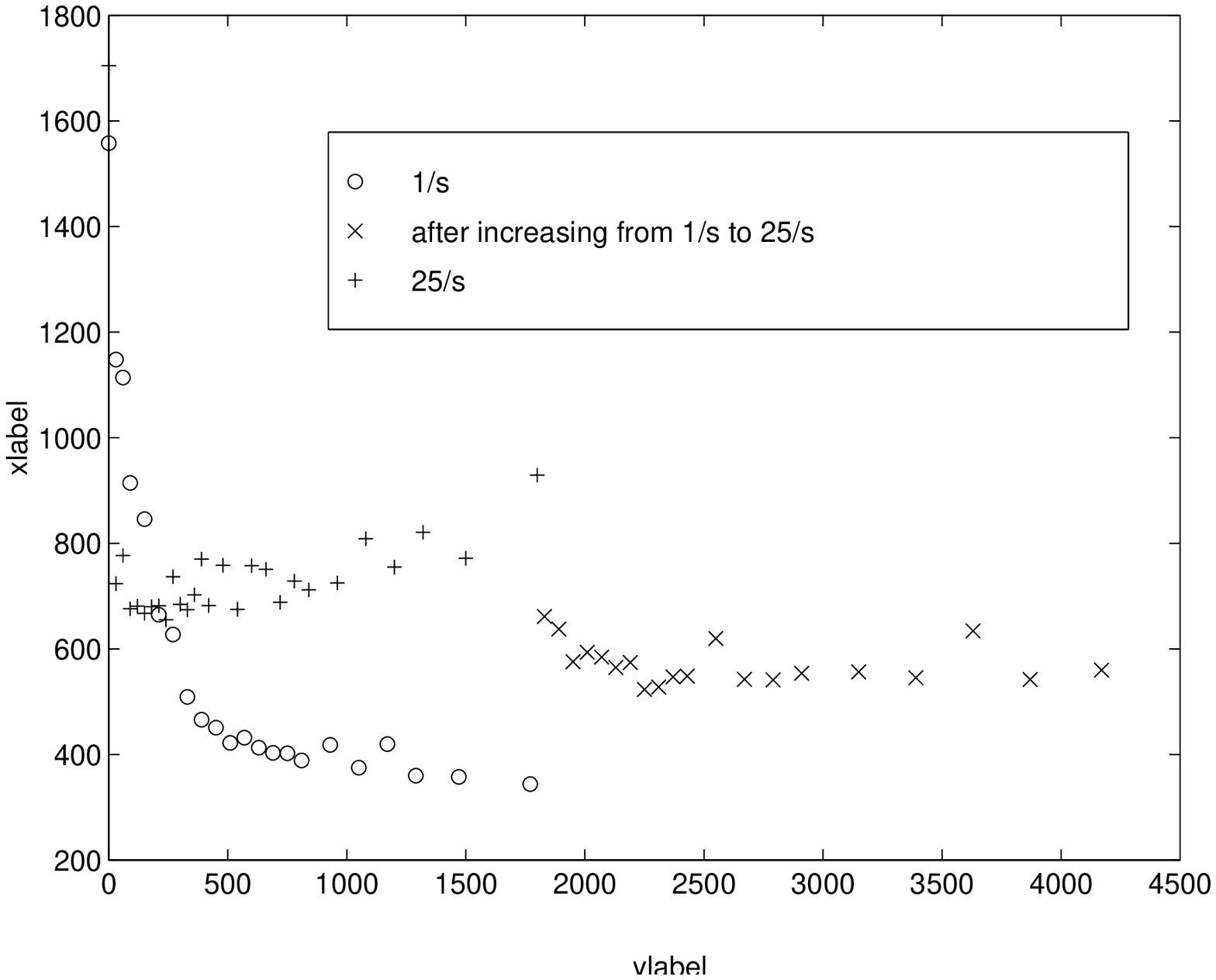,width=11cm,bbllx=90,bblly=246,bburx =540,bbury=560}
\end{center}
\vspace{2cm}
\caption{ $G^{\prime}$  as a function of shear time $t_s$, $f$=1Hz, T=25 $^{\circ}$C. Note the jump in the modulus when $\gs$ was changed abruptly from
1 s$^{-1}$ to 25 s$^{-1}$.}
\label{fig:rheo4}
\end{figure}

In the absence of added particles, we subject the lamellar-phase samples
to steady shear for a time $t_s$, stop and measure the
small-amplitude\footnote{strain amplitude of 5 $\times$ 10$^{-3}$; data
for smaller amplitudes were too noisy} dynamic moduli over a wide
frequency range, resume shearing, and repeat this procedure for a total
shearing time of upto an hour and a half. In some cases, we sheared at 1
s$^{-1}$ for half an hour before increasing the shear-rate to 25
s$^{-1}$, measuring dynamic moduli as before. (While we observed
variation of up to 10\% in the dynamic modulii between samples, the
qualitative features and trends remained unchanged.) The results of this
procedure, summarised in Figs. \ref{fig:rheo1} to \ref{fig:rheo4}, are
as follows: (i) The samples initially had $\gp$, $\gpp$ of about 1500 Pa
and 400 Pa respectively at $\omega = 2 \pi $ rad/sec. At any given
shear-rate (ranging from 1 to 50 s$^{-1}$), the values of $G'$ and $G''$
at fixed $\omega$ show a general tendency to decrease with time,
settling down to a steady-state value at long times, as seen in Figs.\
\ref{fig:rheo1} and \ref{fig:rheo1g2}.\vspace{1cm}
The decrease is not monotone, nor
is the steady state truly steady; there are slow oscillations or even a
tendency for the moduli to increase at long times. We suspect the origin
of the oscillations is in the small sample thickness ($d=100\mu$m), or
in irregularities introduced by flow startup. The long-time increase may
be a result of the working-in of defects by shear. As mentioned in
section III, the video clips do show the formation of a grid-like
pattern in the initially homeotropic region as the sample is being
sheared. This may contribute to the long-time increase in the moduli.
\vspace{1cm}

(ii) The lower the imposed shear-rate, the lower the steady-state values
of $G'$ and $G''$ (at the reference frequency of 2$\pi$ rad/s). (iii)
The overall {\em shapes} of the rheometric spectra $\gp$ and $\gpp$ (Figs.
\ref{fig:rheo2} and \ref{fig:rheo2g2})
remain invariant under shear-treatment, although the magnitudes of the
moduli, as well as the frequency locations of specific features depend
on the duration and shear-rate of the shear-treatment. The $\gpp$
spectrum has roughly a $\omega^{1/2}$ form at large $\omega$ and is
rather flat at small $\omega$, suggesting the presence of a very large
intrinsic time scale in the system \cite{barnes}. The origin of this
long time scale, as in many soft and ill-characterised solids, remains
unclear \cite{kawa,cates,weitzetal}. A similar variation in the
frequency response of $G'$ under shear treatment has been observed in
block copolymers by Riise \etal\ \cite{riise}.
\vspace{1cm}

(iv) The $\gp$ curves for different durations of shear-treatment can be
made to collapse onto the $t_s=0$ (no shear-treatment) curve (Fig.
\ref{fig:rheo3}) if log($\gp/G'_o$) is plotted against
log($\omega/\omega_o$), where $G'_o$ and $\omega_o$ are, respectively, a
reference modulus and a characteristic frequency that shift. We find
that both $G'_o$ and $\omega_o$ decrease monotonically (Fig.
\ref{fig:go}) as $t_s$ increases. This is pleasingly consistent with the
idea of an elasticity arising from a coarsening defect network (see the
discussion in section \ref{summary}). We emphasize that the functions
$G'_o(t_s)$ and $\omega_o(t_s)$ are unique only upto arbitrary
multiplicative factors, as the data collapse can be achieved if the $\gp$
and $\omega$ for no shear-treatment are multiplied by the same
respective factors. For reasons we do not understand, however, the data
on $\gpp$ do not show a satisfactory collapse.
\vspace{1cm}

\vspace{1cm}

(v) After shearing at 1
s$^{-1}$ for about 30 minutes, reaching a quasisteady $G'$ of about 400
Pa, the shear-rate was increased abruptly to 25 s$^{-1}$. It can be seen
from Fig.\ \ref{fig:rheo4} that $G'$ picks up rapidly, showing a
tendency to climb back to the value that it would have had if sheared
directly at 25 s$^{-1}$. Thus, shear on the one hand reduces an
initially high modulus, but is equally capable of increasing an
initially {\em low} modulus. Despite some apparent history-dependence,
we speculate that shear-treatment imposed for a long enough time does
produce a structure with a well-defined dynamic-modulus spectrum
depending only on the value of the shear-rate. We propose to study the
rheological response to small oscillations about the steady flow, not by
stopping the flow, which inevitably introduces artifacts. Linear
response about {\em steady} flow is the true measure of the properties
of the nonequilibrium steady state we are studying, namely the sheared
lamellar phase.
\vspace{1cm}

\begin{figure}
\begin{center}
\epsfig{file=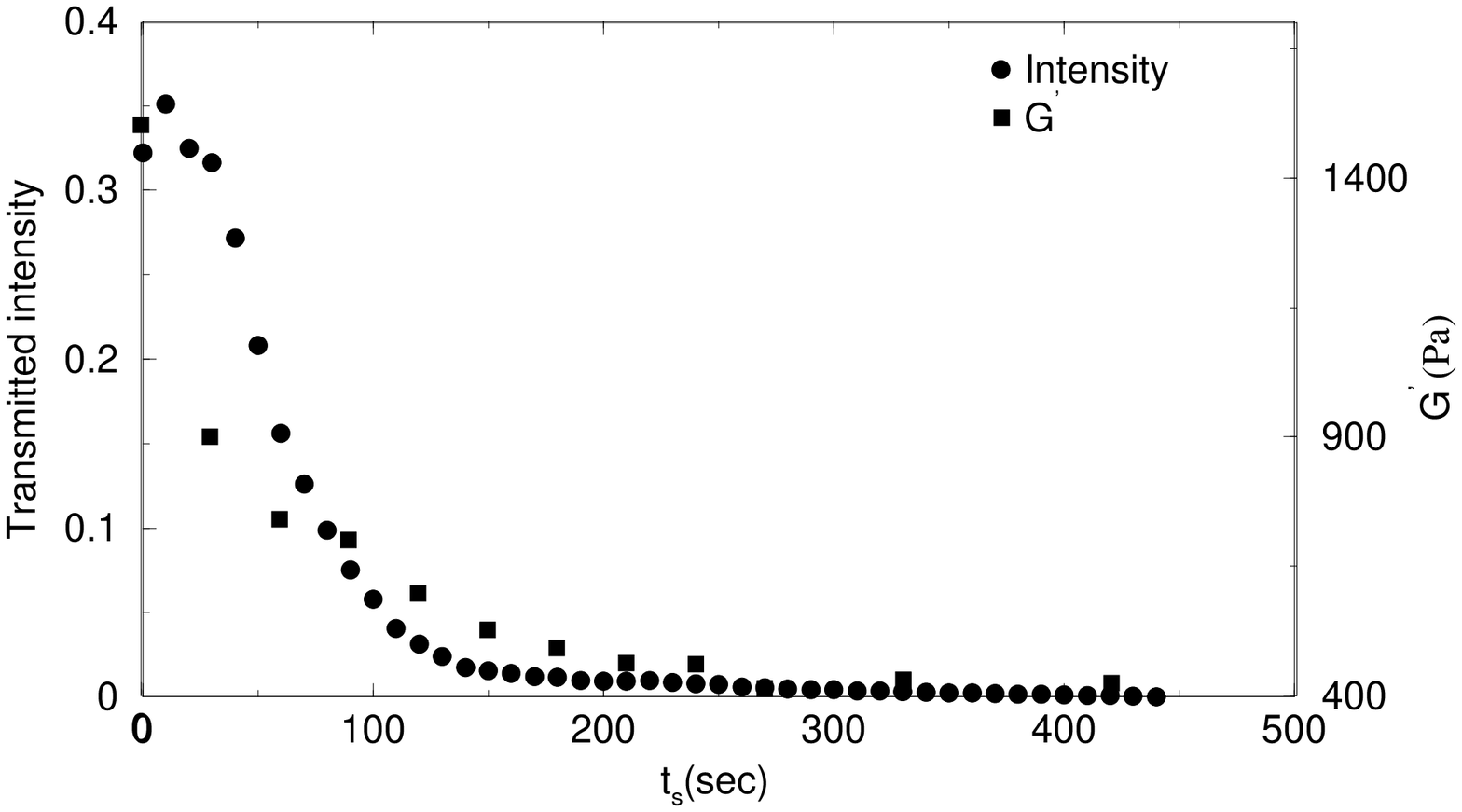, width=10cm}
\end{center}
\caption{ Transmitted intensity and $G'$ of the  lamellar sample
(SLES=73.2 w/w \%) as functions of shear time $t_s$, $\gs$=2 s$^{-1}$,
$d=100 \mu$m.
}
\label{fig:slesig}
\end{figure}

\vspace{2cm}
To compare the rheometry with the light microscopy we have measured the
transmitted light intensity as a function of shear time $t_s$
(Fig.\ \ref{fig:slesig}).
Since homeotropically aligned regions appear dark,
this measures the density of defects. As seen in the figure,
the decay of the transmitted intensity is in clear qualitative agreement
with the decrease in the storage modulus under
similar conditions of shear treatment ($d = 100 \mu$m , $\gs
= 2$ s$^{-1}$), supporting the conjecture that the elasticity arises
primarily from the defect network. A similar correlation has been noted
by Larson \etal \cite{larson2} in thermotropics.
\vspace{1cm}

\vspace{1.5cm}

\subsection{Rheometry with particles}
\label{rheopart}
\begin{figure}
\begin{center}
\psfrag{xlabel}{$t_s$(sec)}
\psfrag{ylabel}{$G'$(Pa)}
\epsfig{file=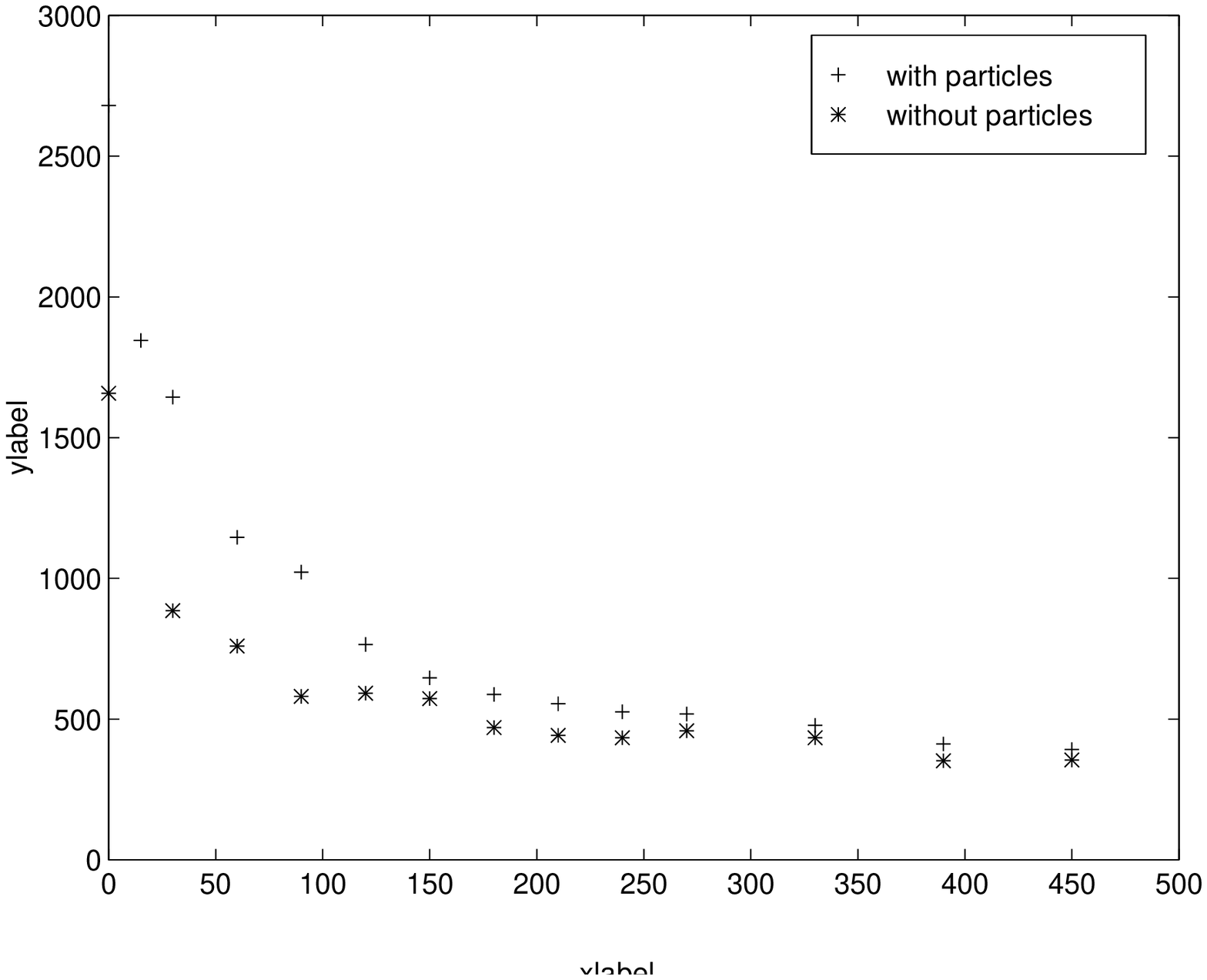,width=11cm,bbllx=90,bblly=246,bburx =540,bbury=560}
\end{center}
\vspace{1.5cm}
\caption{Variation of $G'$ of the lamellar sample (SLES=70w/w \%)
as a function of shear-time $t_s$, with and without particles, $\gs$=2 s$^{-1}$, $f$=1 Hz and T=25 $^\circ$C.
}
\label{fig:g1p10_30}
\end{figure}

\begin{figure}
\begin{center}
\psfrag{xlabel}{$t_s$(sec)}
\psfrag{ylabel}{$G''$(Pa)}
\epsfig{file=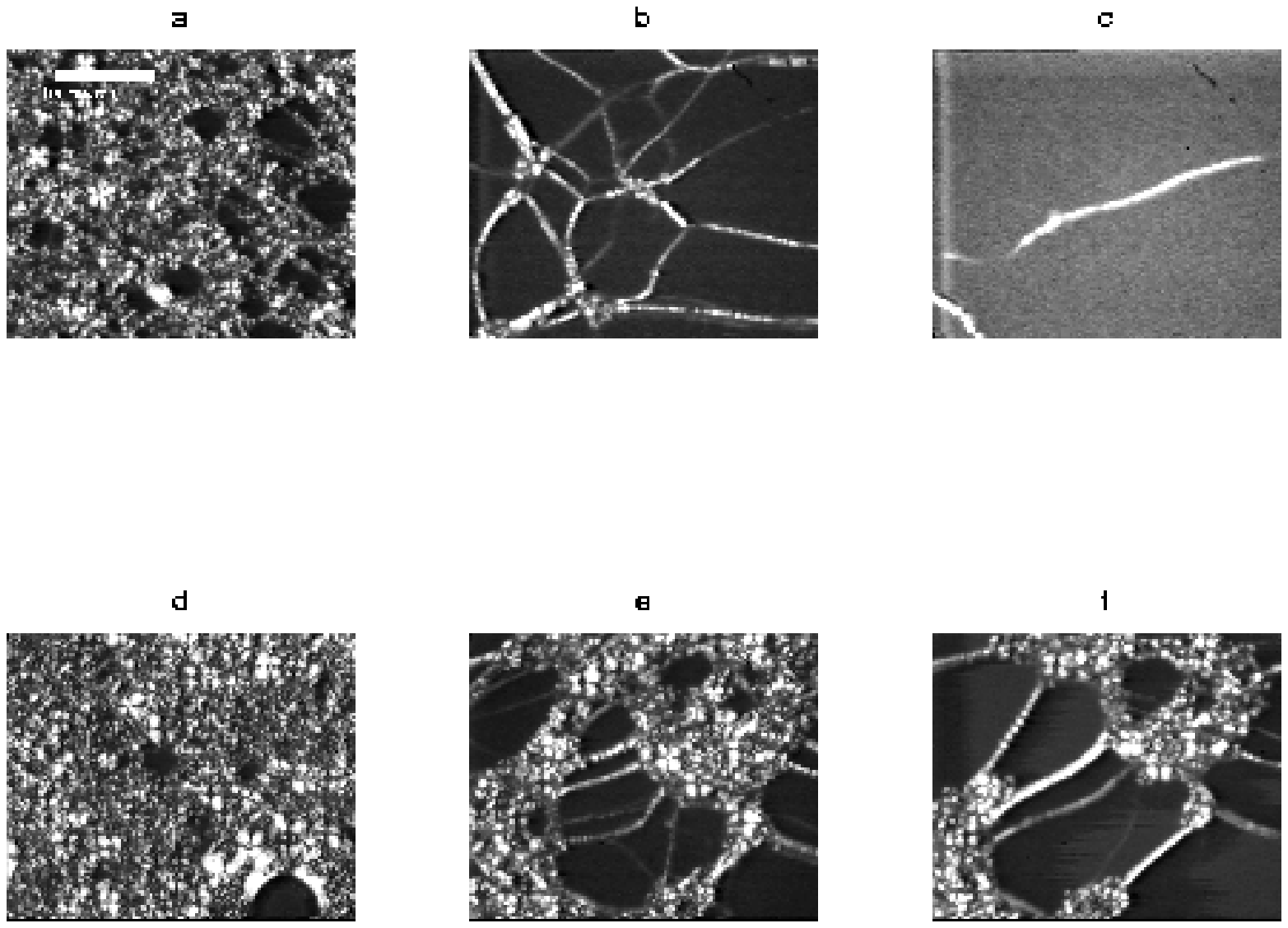,width=11cm,bbllx=90,bblly=246,bburx =540,bbury=560}
\end{center}
\vspace{2cm}
\caption{Variation of $G''$ of the lamellar sample as a function of
shear-time $t_s$, with and without particles,
$\ \gs$=2 s$^{-1}$, $f$=1 Hz and T=25 $^\circ$C.
}
\label{fig:g2p10_30}
\end{figure}

The effect of particulate additives on the viscoelastic properties was
studied by adding monodisperse polystyrene + 30\% polybutylmethacrylate
spheres (9.55 $\pm$ 0.44 $\mu$m diameter, 0.5 \% volume fraction ) to
the lamellar phase sample. The storage modulus is consistently higher in
the presence of particles (Fig.\ \ref{fig:g1p10_30}) but approaches that
of a particle-free sample at the longest times ($\sim$ 10 min). A
similar trend is observed for $G''$ (Fig. \ref{fig:g2p10_30}), although
the data has more scatter. As mentioned in section III, the particles
tend to coagulate at very long times, and the anchoring of the defect
network is lost; this too is in keeping with the ultimate decrease of
the moduli to those of the particle-free sample. The variation in the
dynamic modulii between samples was greater in the presence of particles
than without, roughly 20\%, though part of it could be attributed to the
difficulty in maintaining a constant particle concentration in all
samples, but all the trends were reproducible.
\vspace{1cm}

  \section{Summary and analysis}
\label{summary}
Our video microscopy studies have shown that sheared lyotropic lamellar
phases display a striking defect network structure made up of oily
streaks. These defects anneal away slowly under shear treatment, by a
route which appears to be the thinning and disappearance of lines. The
addition of a small concentration of suspended particulate impurities
greatly retards the decay of this defect network. Previous studies of
particulate additives in a lyotropic gel system showed a large increase
in the rigidity modulus \cite{shouche}. It was conjectured there that
surfactant molecules adsorb on the particles, resulting in direct
interparticle bridges and stress transmission. The oily streak network
we observe is a far more likely candidate for such a stress-transmitting
structure.

We present a rough theoretical estimate of the rigidity of the
network. As in \cite{zap}. we argue that a network of oily-streak lines with
mesh size $\ell$ and tension $\Gamma$ should have a shear modulus
$G' \sim \Gamma / \ell^2$. The value of $\Gamma$ depends on
the internal structure of the oily streaks.
Our images of the oily-streak network (Fig. \ref{fig:images}a, for example,
or other images not presented in the paper)
clearly show the presence of focal domain arrays. We thus estimate the
line tension using the model of \cite{kleman} rather than the
striated dislocation line pairs of \cite{schn}, and find reasonable
agreement with our measurements.
In the focal domain model \cite{kleman}, there are two contributions
to the line energy: a mean-curvature energy piece logarithmic in the width
of the streak, and a major contribution from layer compressions in a
narrow region interpolating between the domains and the bulk undistorted
layering. The curvature term turns out to be negligibly small here,
and we shall ignore it. For an isolated oily streak in a sample of
thickness $h$, the compression energy is found \cite{kleman} to be of
the form
\begin{equation}
\label{streakenergy}
{\cal E} = \bar{B} h^2 \left[{e \over {\sqrt{1 - e^2}}} -
{a \over h} \right]^5
\end{equation}
times a geometrical factor of order unity. Here $\bar{B}$ is the
layer compression modulus at constant chemical
potential of surfactant, and $e$ and $a$ are respectively the eccentricity
and major axis of the ellipses constituting the domains.
In a {\em network} of oily streaks, as distinct from an individual streak,
we expect this dependence on the thickness $h$ to be screened at the scale of
the mesh size $\ell$, i.e., $h$ should be replaced by $\ell$ in
(\ref{streakenergy}). Our observations suggest $e \simeq 0.5$ to $0.6$,
$a \simeq 5$ to $ 10 \mu$m, and $\ell \simeq 50 \mu$m, giving an energy
per unit length ${\cal E} \sim (10^{-3}$ to $10^{-2}) \bar{B} \ell^2$ and hence a
contribution to the shear modulus which scales with but is much smaller in magnitude than the layer compression modulus:
$G' \sim (10^{-3}$ to $10^{-2})\bar{B}$. For $\bar{B} \sim 10^5$ Pa,
this gives $G' \sim 10^2$ to $10^3$ Pa, which is in the right range.
The fifth power in (\ref{streakenergy}) does of course make this estimate
rather sensitive to the precise values of the parameters involved.

While the modulus of an initially defect-ridden state decreases
substantially with shear-treatment, the data for the frequency-dependent
storage modulus after different amounts of shear-treatment can be scaled
onto a single curve, Fig. \ref{fig:rheo3}. This data collapse can be rationalised as
follows: At each stage of the shear-treatment, the network has a
mesh size $\ell$, a characteristic inverse timescale $\omega_o$
(the relaxation rate of structures at the scale $\ell$),
and a characteristic modulus $G'_o$ (the shear modulus of an elementary
cell of the network). It is natural to assume that $\omega_o$
decreases as $\ell$ increases, since a coarser mesh
should relax more slowly, and that $G'_o = \Gamma / \ell^2$,
where $\Gamma$ is the line tension.
This suggests a frequency-dependent shear modulus
$\gp = G'_o F(\omega/\omega_o)$, which should account
for the data collapse provided $\Gamma$ either does
not evolve under coarsening or else evolves in a
manner determined entirely by $\ell$.
The scaling is thus quite easily understood in the focal domains
model \cite{kleman}, in which $\Gamma$ is independent of the
thickness of the lines for a coarse network (see our estimates
in the previous paragraph)\footnote{In the dislocation model \cite{schn,kleman},
however, $\Gamma$ depends on the Burgers-vector content of the line,
i.e., on its thickness, and will thus decrease as $\ell$ increases,
since our oily streaks appear to thin under shear
treatment. We do not consider this case here since our images
show the focal domains of \cite{kleman}.}.

Although the shear modulus of an initially highly rigid (and therefore
presumably very defect-ridden) state decreases gradually with time as
the state is sheared steadily, an initially low-modulus state (produced by
shear-treatment at low rates) becomes more rigid upon shearing at a high
rate. This tells us that the observed steady-state defect structure is
the result of a competition between the working-in and the working-out
of defects by shear. We conjecture that there is a unique structural and
rheological state associated with a given shear-rate rather than a given
shear-history. The data does show some history-dependence, but we
suspect that this is a transient. Indeed, it would be most appropriate
to study the properties of the sheared lamellar phase not by stopping
the flow but rather by treating it as a nonequilibrium steady state, and
measuring its linear rheological response to a small oscillatory
component superposed on the steady shear. We are currently pursuing such
studies.

\section*{Acknowledgement}
\label{ack}
We thank Prof A. K. Sood for extensive access to imaging
facilities, and R. Adhikari for useful discussions.
Partial funding for this project, including a
Project Associateship for GB, came from Unilever Research India.

\end{document}